\documentclass[aps,twocolumn,nofootinbib,superscriptaddress]{revtex4-1}
\usepackage{amssymb,amsmath,amstext,amsfonts,color,ulem,textcomp}
\usepackage{graphicx}

\newcommand{\beq}{\begin{equation}}
\newcommand{\eeq}{\end{equation}}
\newcommand{\bal}{\begin{aligned}}
\newcommand{\eal}{\end{aligned}}

\begin{document}

\title{A theory for multiple partially massless spin-2 fields} 

\author{Nicolas Boulanger}
\affiliation{Service de Physique de l'Univers, Champs et Gravitation, Universit\'e de Mons, UMONS Research Institute for Complex Systems, Place du Parc 20, 7000 Mons, Belgium}
\author{C\'edric Deffayet}
\affiliation{Sorbonne Universit\'e, UPMC Paris 6 and CNRS, UMR 7095, Institut d'Astrophysique de Paris, GReCO, 
98bis Boulevard Arago, 75014 Paris, France}
\affiliation{IHES, Le Bois-Marie, 35 Route de Chartres, 91440 Bures-sur-Yvette, France}
\author{Sebastian Garcia-Saenz}
\affiliation{Sorbonne Universit\'e, UPMC Paris 6 and CNRS, UMR 7095, Institut d'Astrophysique de Paris, GReCO, 
98bis Boulevard Arago, 75014 Paris, France}
\author{Lucas Traina}
\affiliation{Service de Physique de l'Univers, Champs et Gravitation, Universit\'e de Mons, UMONS Research Institute for Complex Systems, Place du Parc 20, 7000 Mons, Belgium}

\begin{abstract}

We revisit the problem of building consistent interactions for a multiplet of partially massless spin-2 fields in (anti-)de Sitter space. After rederiving and strengthening the existing no-go result on the impossibility of Yang--Mills type non-abelian deformations of the partially massless gauge algebra, we prove the uniqueness of the cubic interaction vertex and field-dependent gauge transformation that generalize the structures known from single-field analyses and in four spacetime dimensions, where our results also hold. Unlike in the case of one partially massless field, however, we show that for two or more particle species the cubic deformations can be made consistent at the complete non-linear level, albeit at the expense of allowing for negative relative signs between kinetic terms, making our new theory akin to conformal gravity. Our construction thus provides the first example of an interacting theory containing only partially massless fields.

\end{abstract} 

\maketitle


\section{Introduction}

De Sitter and anti-de Sitter ((A)dS) spaces allow for the existence of exotic particles known as partially massless (PM), which have no counterpart in flat space. PM fields have physical properties intermediate between those of massless and generic massive fields: they possess a non-zero mass proportional to the (A)dS curvature scale, and also a gauge symmetry that reduces the number of propagating degrees of freedom \cite{Deser:1983mm,Deser:2001us}. The simplest PM field has spin $s=2$, and the possible relevance of a PM graviton in the contexts of gravitational physics and cosmology has attracted much attention over the past decade due its potential role in novel approaches to the cosmological constant problem and its distinctive signatures in the physics of the early universe \cite{Baumann:2017jvh,Franciolini:2017ktv,Arkani-Hamed:2018kmz,Goon:2018fyu}. At a more formal level, PM fields are of interest in the endeavor to understand the structure of higher-spin gauge theories for which the irreducible representations of the (A)dS group play a prominent role \cite{Bekaert:2013zya,Brust:2016zns,Brust:2016xif,Basile:2016aen,Garcia-Saenz:2018wnw}.

In view of their theoretical and phenomenological interest, building consistent interacting theories involving PM fields remains a major open problem in field theory. While interactions for PM fields of spin $s>2$ remain largely unexplored, the spin-2 case has proved amenable to some detailed analyses (see e.g.\ \cite{Zinoviev:2006im,deRham:2012kf,Hassan:2012gz,Deser:2013uy,Apolo:2016ort,Apolo:2016vkn}). An important outcome of these studies is the no-go theorem stating the absence of two-derivative interaction vertices for a single PM graviton \cite{deRham:2013wv,Joung:2014aba,Garcia-Saenz:2014cwa}, suggesting that any non-trivial PM spin-2 theory must contain additional fields. In fact, a consistent theory---in the sense that interactions preserve the counting of degrees of freedom---that contains a PM spin-2 particle does exist: conformal gravity, which can be regarded as a non-linear model for a massless and a PM spin-2 field in (A)dS \cite{Maldacena:2011mk,Deser:2012qg}. Conformal gravity can moreover be extended so as to include interactions among several conformal gravitons \cite{Boulanger:2001he,Boulanger:2002bt}, thus leading a non-linear theory for arbitrarily many massless-PM pairs. (The cubic sector of this generalized model has been rediscovered recently in \cite{Joung:2019wwf}.) To the best of our knowledge, these are the only instances of Lagrangian theories coupling PM fields in a non-trivial way. They necessitate massless spin-2 fields in the spectrum, thus still leaving open the question of whether purely PM theories may be written, even in principle.

A natural starting point when attempting a bottom-up construction of interactions involving only PM particles is to consider a multiplet of PM spin-2 fields $h^a_{\mu\nu}\,$, with $a=1,\ldots,N\,$. Indeed, asking the same question in relation to massless spin-1 fields leads one to Yang--Mills theory \cite{Deser:1963zzc}, which overcomes the obstructions encountered by the single-field Maxwell theory, namely the impossibility of generalizing the gauge algebra and gauge symmetry in order to allow for non-trivial interactions. For a collection of PM spin-2 fields the problem was first addressed in \cite{Garcia-Saenz:2015mqi}, where it was established that the gauge algebra of the PM symmetry does not admit any non-abelian extension to first non-trivial order in the fields. This however does not rule out the existence of non-linear deformations of the gauge symmetry or the possibility of constructing interactions. In addition, it does not rule out more general non-abelian deformations of the gauge algebra at higher order in the fields. In the present work we revisit these two questions and give the following answers: {\it (i) given two or more PM spin-2 fields, there exist two-derivative cubic interaction vertices, allowed only in four dimensions, that require a field-dependent deformation of the PM gauge symmetry and are consistent at the fully non-linear level, although with the restriction that some kinetic terms must have negative relative signs; and (ii) there is no possible non-abelian deformation of the gauge algebra---it remains abelian to all orders in perturbations.} As a corollary, in the case where all kinetic terms have the same sign, our results extend the no-go theorems \cite{deRham:2013wv,Joung:2014aba,Garcia-Saenz:2014cwa} on the impossibility of two-derivative interactions to an arbitrary number of PM spin-2 fields.


\section{Deformation analysis}

Our starting point is the action for a collection of $N$ free PM spin-2 fields $h^a_{\mu\nu}$,
\beq \label{eq:free action}
S_0=-\frac{1}{4}\int d^Dx\sqrt{-g}\,k_{ab}\left[F^{a\mu\nu\rho}F^b_{\mu\nu\rho}-2F^{a\mu}F^b_{\mu}\right]\,,
\eeq
where $F^a_{\mu\nu\rho}:=\nabla_{\mu}h^a_{\nu\rho}-\nabla_{\nu}h^a_{\mu\rho}$, $F^a_{\mu}:=g^{\nu\rho}F^a_{\mu\nu\rho}$, and $k_{ab}$ is an internal metric that may be chosen to be a diagonal matrix with entries $+1$ and $-1$.\footnote{Latin or ``color'' indices label the different fields, and they are raised and lowered with the internal metric $k_{ab}$. Greek indices label the coordinates, and they are raised and lowered with the background space metric $g_{\mu\nu}$. We initially work in an arbitrary number $D$ of spacetime dimensions and use the mostly-plus metric signature. Covariant derivatives are compatible with the metric tensor $g$ of the (A)dS background. The curvature scalar of the background is $-\sigma/L^2$, with $\sigma=+1$ for AdS and $\sigma=-1$ for dS.} In a dS background, a unitary theory corresponds to $k_{ab}=\delta_{ab}$, although we will see that this choice does not admit fully consistent cubic interactions. The tensor $F^a_{\mu\nu\rho}$ is an abelian field strength in the sense that it is invariant under the PM gauge symmetry,
\beq \label{eq:free PM symmetry}
\delta_{\epsilon}^{(0)}h^a_{\mu\nu}=\nabla_{\mu}\nabla_{\nu}\epsilon^a-\frac{\sigma}{L^2}\,g_{\mu\nu}\epsilon^a\,.
\eeq

The goal of the deformation analysis is to extend the action \eqref{eq:free action} with non-trivial interactions in a consistent manner, i.e.\ while maintaining the number of gauge symmetries. In general, this may require modifying the gauge transformation law in \eqref{eq:free PM symmetry} with field-dependent terms, schematically $\delta_{\epsilon}=\delta_{\epsilon}^{(0)}+\delta_{\epsilon}^{(1)}+\delta_{\epsilon}^{(2)}+\cdots$, in a way that the non-linear action, $S=S_0+S_1+S_2+\cdots$, respects the gauge invariance, that is $\delta_{\epsilon}S=0$. In our case, since $S_0$ is quadratic in the fields, $S_1$ will encode the cubic interaction vertices, $S_2$ the quartic ones, and so on. In practice we will introduce a bookkeeping parameter $\alpha$ to perform the perturbative expansion, so that $S_n$ and $\delta^{(n)}_{\epsilon}$ are each proportional to $\alpha^n$.

Instead of solving directly for the deformations $S_n$ and $\delta^{(n)}_{\epsilon}$, what is known as the Noether procedure as spelled out in \cite{Berends:1984rq}, we make use its cohomological reformulation \cite{Barnich:1993vg}. This method is specially well suited to deal with ambiguities related to trivial deformations arising from redefinitions of the fields and gauge parameters. We refer the reader to the Section 2 of \cite{Boulanger:2000rq} and to \cite{Henneaux:1997bm} for pedagogical introductions. The case of deformations of massive theories was analyzed in the same framework in \cite{Boulanger:2018dau}.


\section{First-order deformations}

At the first order in the deformation procedure we seek to extend the gauge algebra, gauge symmetry and classical action to leading non-trivial order in the fields.

\subsection{Gauge algebra}

The consistency requirement that gauge transformations must form an algebra can be written as 
\beq \label{eq:general gauge algebra}
\left[\delta_{\epsilon_1}\,,\, \delta_{\epsilon_2}\right]h^a_{\mu\nu}=\delta_{\chi}h^a_{\mu\nu} + \mbox{trivial}\,,
\eeq
where ``trivial'' denotes gauge transformations that vanish on-shell and that leave the action identically invariant, and for some functional $\chi$ that depends on the gauge parameters $\epsilon_1$ and $\epsilon_2$ as well as possibly on the fields. An important advantage of the cohomological approach is that $\chi$, and hence the structure of the algebra, can be strongly constrained from algebraic considerations with no a priori knowledge of the possible deformations of the gauge symmetry itself.

At zeroth order in our deformation parameter $\alpha$ we obviously have $\left[\delta_{\epsilon_1}\,,\,\delta_{\epsilon_2}\right]h^a_{\mu\nu}=0+O(\alpha)$, stating that the algebra of the free theory is abelian. At first order we find the unique candidate extension to be given by $\left[\delta_{\epsilon_1}\,,\, \delta_{\epsilon_2}\right]h^a_{\mu\nu}=\delta^{(0)}_{\chi}h^a_{\mu\nu}+O(\alpha^2)$, with $\delta^{(0)}_{\chi}$ as in Eq.\ \eqref{eq:free PM symmetry} and
\beq
\chi=\alpha\left(m^a_{\phantom{a}bc}\epsilon_1^b\epsilon_2^c+n^a_{\phantom{a}bc}\nabla^{\mu}\epsilon_1^b\nabla_{\mu}\epsilon_2^c\right)+O(\alpha^2)\,,
\eeq
where $m^a_{\phantom{a}bc}=m^a_{\phantom{a}[bc]}$ and $n^a_{\phantom{a}bc}=n^a_{\phantom{a}[bc]}$ are otherwise arbitrary at this stage. They correspond to the structure constants of the gauge algebra. In fact, it can be shown that there are no terms of order $\alpha^2$ in the above expression for $\chi\,$. That $\chi$ is field-independent is not an assumption and can be proved indeed. The proof is rather technical and will be presented elsewhere, but it is similar to the corresponding proof given in Section 7 of \cite{Boulanger:2000rq} for the case of massless spin-2 fields and combines it with embedding-space techniques as used e.g.\ in \cite{Bekaert:2010hk,Joung:2014aba,Bonifacio:2018zex}.

Further constraints on $m^a_{\phantom{a}bc}$ and $n^a_{\phantom{a}bc}$ arise by demanding that the algebra is realized on the fields by some infinitesimal gauge symmetry. We find that this requirement is very strong and leads to the result
\beq
m^a_{\phantom{a}bc}=0\,,\qquad n^a_{\phantom{a}bc}=0\,.
\eeq
This implies that the PM spin-2 gauge algebra does not allow for any non-abelian extension. This no-go result was first established in \cite{Garcia-Saenz:2015mqi}, although the present derivation is stronger in that no assumption is needed on the number of derivatives entering in the algebra or on the (in)dependence of $\chi$ on the fields---our result remains true to all orders in perturbations.

\subsection{Gauge symmetry}

An abelian gauge algebra does not imply the absence of non-trivial extensions of the gauge symmetry. For instance, the Chapline--Manton and Freedman--Townsend theories of differential form fields belong to this class \cite{Freedman:1980us,Chapline:1982ww,Henneaux:1997ha}. In the present setting, abelian deformations of the PM gauge symmetry are simple to classify, since they are constructed solely out of the field strength tensor $F^a_{\mu\nu\rho}$ and its derivatives. At first order we restrict our attention to contractions that are linear in $F^a_{\mu\nu\rho}$, finding the following six candidate structures:
\beq\bal \label{eq:general a1bar}
\delta^{(1)}_{\epsilon}h^a_{\mu\nu}&=\alpha\Big(u^{\;\;a}_{(1)bc}\nabla^{\rho}F^b_{\rho(\mu\nu)}\epsilon^c+u^{\;\;a}_{(2)bc}\nabla_{(\mu}F^b_{\nu)}\epsilon^c\\
&\quad +u^{\;\;a}_{(3)bc}\,g_{\mu\nu}\nabla^{\rho}F^b_{\rho}\epsilon^c+v^{\;\;a}_{(1)bc}F^b_{\rho(\mu\nu)}\nabla^{\rho}\epsilon^c\\
&\quad +v^{\;\;a}_{(2)bc}F^b_{(\mu}\nabla_{\nu)}\epsilon^c+v^{\;\;a}_{(3)bc}\,g_{\mu\nu}F^b_{\rho}\nabla^{\rho}\epsilon^c\Big)\,,
\eal\eeq
where the constants $u^{\;\;a}_{(i)bc}$ and $v^{\;\;a}_{(i)bc}$ are arbitrary at this stage of the calculation. We remark that this ansatz is the most general one containing two derivatives. Although this is a restrictive assumption, it is enough for our purposes as it will allow us to classify all cubic interaction vertices that have no more than two derivatives. Notice that, a priori, gauge transformation terms bringing more than two derivatives could be required in order to produce a vertex with two derivatives or less. However, we checked that this is not the case for the couplings of PM spin-2 fields.

\subsection{Cubic action}

Although the result in Eq.\ \eqref{eq:general a1bar} is consistent from the point of view of the gauge algebra, there is of course no guarantee that there exists a local action that realizes this symmetry in full. By demanding consistency with the existence of non-trivial cubic interactions with no more than two derivatives we obtain that the constants $u^{\;\;a}_{(i)bc}$ and $v^{\;\;a}_{(i)bc}$ are all forced to vanish with the exception of $v^{\;\;a}_{(1)bc}\equiv f^a_{\phantom{a}b,c}$, that is
\beq \label{eq:consistent gauge sym}
\delta^{(1)}_{\epsilon}h^a_{\mu\nu}=\alpha\,f^a_{\phantom{a}b,c}F^b_{\rho(\mu\nu)}\nabla^{\rho}\epsilon^c\,,
\eeq
and $f_{ab,c}$ must be symmetric under the exchange of the first two indices. Moreover, this non-trivial possibility is only available when the spacetime dimension is $D=4$. The cubic vertex is given explicitly by
\beq \label{eq:cubic vertex}
S_1=\alpha\int d^4x\sqrt{-g}\,h^a_{\mu\nu}J^{\mu\nu}_a\,,
\eeq
where
\beq\bal
J^{\mu\nu}_a&:=f_{bc,a}\bigg[F^{b\mu}_{\phantom{b\mu}\rho\sigma}F^{c\nu\rho\sigma}-F^{b\mu}F^{c\nu}-F^{b\rho(\mu\nu)}F^c_{\rho}\\
&\quad -\frac{1}{4}\,g^{\mu\nu}F^{b\rho\sigma\lambda}F^c_{\rho\sigma\lambda}+\frac{1}{2}\,g^{\mu\nu}F^{b\rho}F^c_{\rho}\bigg]\,.
\eal\eeq
Given the symmetries of the constants $f_{ab,c}$, we have that the number of independent non-trivial deformations of the free PM spin-2 theory is given by $\frac{1}{2}N^2(N+1)$ at this order in the analysis.

The consistency of the cubic action \eqref{eq:cubic vertex} is rather easy to check (the strength of our result lies in having proved its uniqueness): given that $J^{\mu\nu}_a$ is manifestly invariant under the undeformed PM symmetry \eqref{eq:free PM symmetry}, it suffices to observe that it defines a conserved current in the sense that
\beq
\nabla_{\mu}\nabla_{\nu}J^{\mu\nu}_a-\frac{\sigma}{L^2}\,g_{\mu\nu}J^{\mu\nu}_a\approx 0\,,
\eeq
where ``$\approx$'' means equality modulo the equations of motion of the free theory.

The current $J^{\mu\nu}_a$ actually satisfies stronger conditions: it is identically traceless in $D=4$, i.e.\ $g_{\mu\nu}J^{\mu\nu}_a=0\,$, and it is (covariantly) conserved in the usual sense, i.e.\ $\nabla_{\nu}J^{\mu\nu}_a\approx0\,$. These properties stem from the fact that $J^{\mu\nu}_a$ is related to the Noether current associated with some rigid symmetries of the free PM theory. Explicitly, defining
\beq \label{eq:noether current}
{\cal J}^{\mu}_{ab}:=\sqrt{-g}\,J^{\mu\nu}_a\nabla_{\nu}\bar{\epsilon}_b\,,
\eeq
it is straightforward to verify that ${\cal J}^{\mu}_{ab}$ is a true Noether current in the sense that $\partial_{\mu}{\cal J}^{\mu}_{ab}\approx0\,$, and again only in $D=4$ dimensions. The function $\bar{\epsilon}_a$ in Eq.\ \eqref{eq:noether current} is by definition a Killing parameter of the free theory, i.e., a solution of $\nabla_{\mu}\nabla_{\nu}\bar{\epsilon}_a-\frac{\sigma}{L^2}\,g_{\mu\nu}\bar{\epsilon}_a=0$.\footnote{Explicit expressions for the Killing parameters associated to the PM spin-2 theory have been found in \cite{Hinterbichler:2015nua} for $D=4$ dimensions (although the procedure may be readily generalized to arbitrary $D$).} The corresponding rigid symmetry of the quadratic theory is obtained by considering \eqref{eq:consistent gauge sym} where the gauge parameters $\epsilon^a$ are replaced by $\bar\epsilon^a\,$.


\section{Higher-order consistency}

Having found the most general first-order deformation of the PM spin-2 gauge symmetry and classical action (assuming up to two-derivative interactions), we now turn to the question of its consistency at higher orders in perturbations.

\subsection{Consistency of the deformed gauge symmetry}

The statement that a gauge symmetry must be consistent with an algebra leads to further constraints at higher orders in the deformation analysis. For instance, in Yang--Mills theory, the consistency of the extended gauge transformation law implies the Jacobi identity on the structure constants \cite{Wald:1986bj,Barnich:1994mt}. A similar quadratic constraint also applies to the coefficients of the first-order extension of the PM gauge symmetry that we derived in the previous section, Eq.\ \eqref{eq:consistent gauge sym}. We find that the constants $f_{ab,c}$ must satisfy
\beq \label{eq:first quadratic constraint}
f_{ae,b}f^{e}_{\phantom{e}c,d}\equiv k^{ef}f_{ea,b}f_{fc,d}=0\,.
\eeq
Two simple conclusions readily follow. The first is that in the case of one field ($N=1$) one immediately gets that $f_{11,1}=0$, implying the failure of the field-dependent PM gauge symmetry to extend beyond lowest order. This is the well known no-go result on the absence of deformations for a single PM spin-2 field.

The second remark is that solutions to \eqref{eq:first quadratic constraint} do not exist when $k_{ab}=\delta_{ab}$. Indeed, if this were the case, taking $c=a$ and $d=b$ (with no summation) in Eq.\ \eqref{eq:first quadratic constraint} leads to the conclusion that $f_{ea,b}=0$. It follows that the consistency of the deformed gauge symmetry can only be achieved provided that at least one or more of the fields enter in the action with a ``wrong-sign'' kinetic term. Thus we conclude that any non-trivial theory (subject to our assumptions) of multiple PM spin-2 particles must be non-unitary. As a corollary, for a positive-definite internal metric our result provides a rigorous extension of the aforementioned no-go result to multiple PM spin-2 fields.

\subsection{Consistency of the deformed action}

The consistency of the deformed classical action is simply the requirement of gauge invariance at higher orders in perturbations. In our setting, this is the statement that the extended action $S_0+S_1$ be invariant under the deformed symmetry $\delta^{(0)}_{\epsilon}+\delta^{(1)}_{\epsilon}$. We find
\beq\bal \label{eq:obstruction invariance}
&(\delta^{(0)}_{\epsilon}+\delta^{(1)}_{\epsilon})(S_0+S_1)= \\
& 2\alpha^2f_{ab,e}f^e_{\phantom{e}c,d}\int d^4x\sqrt{-g}\Big[F^{a\mu}_{\phantom{a\mu}\rho\sigma}F^{b\nu\rho\sigma}F^c_{\lambda\mu\nu}\nabla^{\lambda}\epsilon^d+\cdots\Big]\,,
\eal\eeq
where the ellipses contain terms that vanish on the free equations of motion, and which may be removed by extending the gauge symmetry with appropriate $O(\alpha^2)$ terms. On the other hand, the expression shown represents an obstruction to the invariance of the action that may in principle be removed by the inclusion of quartic interactions to the action, i.e.\ by a new deformation term $S_2$. Although it would be interesting to pursue this route, it goes beyond our present scope as it would require a cumbersome classification of candidate quartic operators as well as consistency checks at $O(\alpha^3)$.

We will instead impose the vanishing of the obstruction in Eq.\ \eqref{eq:obstruction invariance} by choosing coefficients $f_{ab,c}$ satisfying the second quadratic constraint
\beq \label{eq:second quadratic constraint}
f_{ab,e}f^{e}_{\phantom{e}c,d}\equiv k^{ef}f_{ab,e}f_{fc,d}=0\,.
\eeq
In conclusion, the unique first-order deformations to the PM gauge symmetry (Eq.\ \eqref{eq:consistent gauge sym}) and PM action (Eq.\ \eqref{eq:cubic vertex}) remain consistent at the complete non-linear level provided (a) that no further deformations are introduced beyond $O(\alpha)$, and (b) that the constants $f_{ab,c}$ satisfy the constraints in Eqs.\ \eqref{eq:first quadratic constraint} and \eqref{eq:second quadratic constraint}.

Understanding the space of solutions of the quadratic constraints \eqref{eq:first quadratic constraint} and \eqref{eq:second quadratic constraint} (which depends of course on the signature of the internal metric $k_{ab}$ and on the number $N$ of particle species) is a highly non-trivial task that will be addressed elsewhere. However we can show that there is a unique solution (modulo trivial rescalings) for $N=2$ and $k_{ab}={\rm diag}(+1,-1)$, and have found particular solutions for all $N\geq3$ and different choices of $k_{ab}$ that give rise to three-particle interactions. They are given explicitly in the Appendix.


\section{Discussion}

In the Introduction we recalled the existence of the fully nonlinear theories of multi-conformal gravitons obtained in \cite{Boulanger:2001he,Boulanger:2002bt}, corresponding to models that couple a set of massless spin-2 fields to the same number of PM spin-2 fields. Given that the family of theories we have uncovered here in four dimensions are built on the cubic vertex of conformal gravity, this shows a posteriori that there does exist a fully nonlinear truncation of the conformal multi-graviton models of \cite{Boulanger:2001he,Boulanger:2002bt} to an interacting PM theory. This truncation should be done in conjunction with the imposition of the constraint that the internal structure constants $a^a_{\phantom{a}bc}$ of the multi-conformal graviton theories should satisfy our quadratic constraints \eqref{eq:first quadratic constraint} and \eqref{eq:second quadratic constraint}. This consistent truncation thereby avoids the no-go results obtained in \cite{Deser:2013bs,Joung:2014aba}, thanks to our allowing for several conformal gravitons from the beginning.

More interestingly, our findings show that there exist \textit{other} branches of consistent theories of interacting PM fields, that \textit{cannot} be attained by a truncation of the multi-conformal graviton models. This happens when the structure constants $f_{ab,c}$ entering our model are not totally symmetric. Indeed, the aforementioned theories of conformal multi-gravity necessitate completely symmetric internal structure  constants $a_{abc}=a_{(abc)}\,$, while the present construction allows for coupling constants $f_{ab,c}$ which are not. One such solution with a mixed-symmetric $f_{ab,c}$ is given in the Appendix.

Our results are relevant in that they provide a proof of principle that it is possible to construct non-trivial theories of finitely many PM fields that are fully consistent from the point of view of the gauge structure. It is indeed an outstanding field theoretical problem to determine what theories containing PM fields may in principle be written purely on the basis of consistency, and our construction is a step forward in this program. The requirement that some fields must have negative kinetic energy around (A)dS space could in principle be evaded by changing the background to a less symmetric one or by embedding our model in a theory with dynamical gravity. It is also feasible that our non-linear theory (or a generalization thereof) may allow for non-trivial solutions about which fluctuations are no longer ghostly, a phenomenon that is known to occur for instance in massive gravity \cite{Gabadadze:2003jq}. We hope to return to these issues in the future. Regarding the applicability of our model of multiple PM spin-2 particles, it would be interesting to see whether it could be embedded into a higher-spin theory that should in turn provide an extension of the higher-spin model proposed in \cite{Bekaert:2013zya} and discussed later in \cite{Brust:2016zns,Brust:2016xif}. This theory is of interest as it has been conjectured to be dual to the $O(N)$ model at a multi-critical isotropic Lifshitz point (see \cite{Diehl} for a review). The non-unitarity of the dual theory is not necessarily a pathology in this context, as non-unitary conformal theories are of potential interest in condensed matter physics \cite{Maassarani:1996jn} (see also e.g.\ \cite{ElShowk:2011gz,Penedones:2015aga} for other applications of non-unitary CFTs). It could actually be more likely, in connection to what we explained previously, that our model could be embedded into an extended version of conformal higher-spin gravity (see e.g.\ \cite{Adamo:2018srx,Basile:2018eac}), the existence of which is an interesting question on its own.

 
\begin{acknowledgments}

We would like to thank Andrea Campoleoni, Kurt Hinterbichler and Per Sundell for useful comments and discussions. The work of NB is partially supported by an F.R.S.-FNRS PDR grant ``Fundamental issues in extended gravity" \textnumero~T.0022.19. LT is Research Fellow of the Fund for Scientific Research-FNRS. He wants to thank IAP for kind hospitality. CD and SGS are supported by the European Research Council under the European Community's Seventh Framework Programme (FP7/2007-2013 Grant Agreement no.\ 307934, NIRG project). SGS would also like to thank the unit ``Physics of the Universe, Fields and Gravitation" of UMONS for generous hospitality.

\end{acknowledgments}

\appendix

\section{Explicit solutions of quadratic constraints}

{\it Totally symmetric solutions.---}An explicit solution of the quadratic constraints, Eqs.\ \eqref{eq:first quadratic constraint} and \eqref{eq:second quadratic constraint}, is given by
\beq \label{eq:explicit solution}
f_{ab,c}=(N-1)^{(n_g)_{abc}/2}\,,
\eeq
for the choice of internal metric $k_{ab}={\rm diag}(+1,\ldots,+1,-1)\,$, and where $(n_g)_{abc}\in\{0,1,2,3\}$ denotes the number of times that the index ``$N$'' (corresponding to the ``ghostly'' field in our convention) appears in $f_{ab,c}\,$; for instance $f_{NN,N}=(N-1)^{3/2}\,$.

Another simple particular solution that is valid for all even $N$ is $f_{ab,c}=1\,,\;\forall\,a,b,c\in\{1, \dots, N\}\,$, with metric $k_{ab}={\rm diag}(+1,\ldots,+1,-1,\ldots,-1)$ that has the same number of ``$+1$'' and ``$-1$'' in its entries. Notice that both these solutions give a totally symmetric $f_{ab,c}\,$, in which case the two constraints \eqref{eq:first quadratic constraint} and \eqref{eq:second quadratic constraint} are in fact equivalent.

For $N=2$ fields these two solutions reduce to $f_{ab,c}=1\,,\;\forall\,a,b,c\in\{1,2\}\,$, with metric $k_{ab}={\rm diag}(+1,-1)$. In this case we can moreover show that this solution is unique modulo rescalings of the fields and gauge parameters. We remark that for $N\geq3$ the constants \eqref{eq:explicit solution} lead to cubic vertices that couple three distinct fields, so that it is not a trivial extension of the $N=2$ solution.

\vspace{0.5cm}

{\it Mixed symmetric solutions.---}For $N=2$ the unique solution to the constraints was totally symmetric under the exchange of the three indices. However, for $N\geq 3\,$, there also exist solutions for mixed-symmetric constants $f_{ab,c}$. For example, when $N=3$ and the metric is $k_{ab}={\rm diag}(+1,+1,-1)\,$, one such solution is given by
\begin{equation} \label{eq:nonsymsol}
\begin{aligned}
f_{11,1} &= f_{22,2} = 1\,,\\
f_{11,2} &= -f_{12,1} = f_{22,1} = - f_{12,2} = 1\,,\\ 
f_{13,1} & = f_{13,2} = - f_{23,1} = - f_{23,2} = \sqrt{2}\,,\\
f_{33,1} & = f_{33,2} = 2\,, \\
f_{ab,3} & = 0\quad \forall\, a,b \in \{1,2,3\}\,.
\end{aligned}
\end{equation}

$\;$


\bibliographystyle{apsrev4-1}
\bibliography{ColoredPMBiblio}

\end{document}